# Relativistic Double Barrier Problem with Three Sub-Barrier Transmission Resonance Regions

A. D. Alhaidari[a], H. Bahlouli[a,b] and A. Jellal[a,c,d]

[a] *Saudi Center for Theoretical Physics, Dhahran, Saudi Arabia*
[b] *Physics Department, King Fahd University of Petroleum & Minerals, Dhahran 31261, Saudi Arabia*
[c] *Physics Department, King Faisal University, Al-Ahsaa, Saudi Arabia*
[d] *Theoretical Physics Group, Faculty of Science, Chouaib Doukkali University, 24000 El Jadida, Morocco*

We obtain exact scattering solutions of the Dirac equation in 1+1 dimensions for a double square barrier vector potential. The potential floor between the two barriers is higher than $2mc^2$ whereas the top of the barriers is at least $2mc^2$ above the floor. The relativistic version of the conventional double barrier transmission resonance is obtained for energies within $\pm mc^2$ from the height of the barriers. However, we also find two more (sub-barrier) transmission resonance regions below the conventional one. Both are located within the two Klein energy zones and characterized by resonances that are broader than the conventional ones.

## I. INTRODUCTION

The basic equation of relativistic quantum mechanics was formulated more than 80 years ago by Paul Dirac [1]. It describes the state of electrons in a way consistent with special relativity, requiring that electrons have spin $\frac{1}{2}$ and predicting the existence of an antiparticle partner to the electron (the positron). The physics and mathematics of the Dirac equation is very rich, illuminating and provides a theoretical framework for different physical phenomena that are not present in the nonrelativistic regime such as the Klein paradox, super-criticality [1-3] and the quantum Hall effect in graphene [4,5]. It is well known that the Dirac equation has positive as well as negative energy solutions [1]. The positive and negative energy subspaces are completely disconnected. This is a general feature of the solution space of the Dirac equation, which is sometimes overlooked. Since the equation is linear, then the complete solution must be a linear combination of the two. Recently one of the authors (ADA) gave a new approach to the resolution of the famous Klein paradox within relativistic quantum mechanics [6]. This was accomplished by incorporating the missing part of the negative energy solution, which is not taken into account in the traditional solution leading to the correct physical and mathematical interpretations of this phenomenon.

On the other hand, tunneling phenomena played an important role in non relativistic quantum mechanics due to its important application in electronic devices [7]. It was Leo Esaki who discovered a characteristic called negative differential resistance (NDR) whereby, for PN junction diodes, the current voltage characteristics has a sharp peak at a certain voltage associated with resonant tunneling. This constituted the first important confirmation that this phenomenon is due to the quantum mechanical tunneling effect of electrons [8]. Tunneling is a purely quantum phenomenon that happens in the classically forbidden region; its experimental observation constituted a very important support to the quantum theory. On the other hand, the study of tunneling of relativistic particles through one-dimensional potentials has been restricted to some simple configurations such as $\delta$-potentials and square barriers, mainly in the study of the possible relativistic corrections to mesoscopic conduction [9] and the analysis of resonant tunneling through multi-barrier systems [10].

Very recently, electron transport through electrostatic barriers in single and bi-layer graphene has been studied using the Dirac equation and barrier penetration effects analogous to the Klein paradox were noted [11]. The study of transmission resonances in relativistic wave equations in external potentials has been discussed extensively in the literature [12-13]. In this case, for given values of the energy and shape of the barrier, the probability of transmission reaches unity even if the potential strength is larger than the energy of the particle, a phenomenon that is not present in the nonrelativistic case. The relation between low momentum resonances and super-criticality has been established by Dombey *et al.* [3] and Kennedy [14]. Some results on the scattering of Dirac particles by a one-dimensional potential exhibiting resonant behavior have also been reported [13-15].

Here we are interested in studying the resonant transmission of a beam of relativistic particles by two separated square barriers with elevated in-between floor and investigating transmission resonance in this structure [1]. Under certain conditions, we demonstrate the occurrence of three (sub-barrier) regions of transmission resonances. One of them is the relativistic extension of the conventional nonrelativistic double barrier transmission resonances for energies within $\pm mc^2$ from the height of the barriers. The other two are located within the two Klein energy zones where only positive and negative energy oscillatory solutions coexist at the same energy. The latter resonances are broader than the conventional ones.

## II. SCATTERING SOLUTION OF THE DIRAC EQUATION

The physical configuration associated with the double barrier problem in our study is shown in Figure 1. In the relativistic units $\hbar = c = 1$, the one-dimensional stationary Dirac equation with vector potential coupling can be written as [1]

$$\begin{pmatrix} m+V(x)-E & -\frac{d}{dx} \\ +\frac{d}{dx} & -m+V(x)-E \end{pmatrix} \begin{pmatrix} \psi^+(x) \\ \psi^-(x) \end{pmatrix} = 0 \qquad (1)$$

where $V(x)$ is the time component of the vector potential whose space component vanishes (i.e., gauged away due to gauge invariance). The potential $V(x)$ is defined by

$$V(x) = \begin{cases} 0 & |x| \geq a_+ + a_- \\ V_+ & a_- < |x| < a_+ + a_- \\ V_- & |x| \leq a_- \end{cases} \qquad (2)$$

where $V_\pm$ and $a_\pm$ are positive potential parameters (see Fig. 1) such that $V_- > 2m$ and $V_+ > V_- + 2m$. We divide configuration space according to the piece-wise constant potential sections into three regions numbered 0 and $\pm$ corresponding to $V = 0$ and $V = V_\pm$, respectively. In regions 0, where the potential vanishes, the equation becomes the free Dirac equation that relates the two spinor components as follows

$$\psi^{\mp}(x) = \frac{1}{m \pm E} \frac{d}{dx} \psi^{\pm}(x). \tag{3}$$

This relationship is valid for $E \neq \mp m$. Since the problem is linear and because $E = \mp m$ belongs to the $\mp$tive energy spectrum, then Eq. (3) with the top/bottom sign is valid only for positive/negative energy, respectively. After choosing a sign in Eq. (3) then the other spinor component obeys the following Schrödinger-like second order differential equation

$$\left( \frac{d^2}{dx^2} + E^2 - m^2 \right) \psi^{\pm}(x) = 0. \tag{4}$$

We should emphasize that *Eq. (4) does not give the two components of the spinor that belong to the same energy subspace*. One has to choose one sign in Eq. (4) to obtain only one of the two components then substitute that into Eq. (3) with the corresponding sign to obtain the other component. Now, within the double barrier (the $V_{\pm}$ regions) the same analysis follows but with the substitution $E \to E - V_{\pm}$ giving

$$\psi^{\mp}(x) = \frac{1}{m \pm (E-V)} \frac{d}{dx} \psi^{\pm}(x), \tag{5a}$$

$$\left[ \frac{d^2}{dx^2} + (E-V)^2 - m^2 \right] \psi^{\pm}(x) = 0, \tag{5b}$$

where $V$ stands for either of the two potentials $V_{\pm}$. Generally, in any region of constant potential $V$, positive/negative energy solutions occur for relativistic energies larger/smaller than $V$. Of these, the oscillatory solutions of the form $e^{\pm ikx}$ are for $|E-V| > mc^2$, where $k^2 = |(E-V)^2 - m^2|$. On the other hand, the exponential solutions of the form $e^{\pm kx}$ hold for $|E-V| < mc^2$.

The scattering solution, which is the subject of this work, is for energies $E > m$. It is straightforward to write down the positive and negative energy solutions of Eqs. (3-5). First, we write the wave vector associated with regions of space in which the potential equals to zero, $V_+$, and $V_-$, as

$$k_{\mu}(E) = \sqrt{m^2 - (E - U_{\mu})^2}, \tag{6}$$

where $\mu = 0, +, -$ and $U_{\mu} = \{0, V_+, V_-\}$. This results in oscillatory solutions if $k_{\mu}$ is pure imaginary which happens when $U_{\mu} - m > E > U_{\mu} + m$ (i.e., in the two grey regions of Fig. 1). Otherwise, these solutions are exponentials (i.e., in the white areas of the figure). The oscillatory positive/negative energy solutions are located in the light/dark grey areas of Fig. 1, respectively. We divide configuration space from left to right into five regions indexed by $v = 1, 2, ..., 5$. The general positive energy solution in these regions (both oscillatory and exponentials) can be written as

$$\psi_{\mu,v}(x) = \frac{A_v}{\sqrt{1+|\alpha_{\mu}|^2}} \begin{pmatrix} 1 \\ \alpha_{\mu} \end{pmatrix} e^{k_{\mu}x} + \frac{B_v}{\sqrt{1+|\alpha_{\mu}|^2}} \begin{pmatrix} 1 \\ -\alpha_{\mu} \end{pmatrix} e^{-k_{\mu}x}, \tag{7a}$$

whereas the negative energy solutions are of the form

$$\psi_{\mu,v}(x) = \frac{A_v}{\sqrt{1+|\beta_{\mu}|^2}} \begin{pmatrix} -\beta_{\mu} \\ 1 \end{pmatrix} e^{-k_{\mu}x} + \frac{B_v}{\sqrt{1+|\beta_{\mu}|^2}} \begin{pmatrix} \beta_{\mu} \\ 1 \end{pmatrix} e^{k_{\mu}x}. \tag{7b}$$

$A_v$ and $B_v$ are constants (the complex amplitudes) associated with right and left "traveling" solutions in the $v$-th region, respectively. The energy parameters $\alpha_{\mu}$ and $\beta_{\mu}$ are defined by

$$\alpha_\mu = \sqrt{(m - E + U_\mu)/(m + E - U_\mu)}, \tag{8a}$$

$$\beta_\mu = \sqrt{(m + E - U_\mu)/(m - E + U_\mu)}. \tag{8b}$$

Note that $\beta_\mu = \pm 1/\alpha_\mu$ for real/imaginary values, respectively. The complex constant amplitudes $\{A_\nu, B_\nu\}$ will be determined by the boundary conditions. We should note that the oscillatory solutions, $e^{\pm ikx}$, in Eqs. (7) represent a wave traveling in the $\pm x$ direction for positive energy solutions and in the $\mp x$ direction for negative energy solutions. The solution of the Dirac equation to the right of the double barrier consists of positive energy plane-wave solutions traveling in the $\pm x$ directions. However, the physical boundary conditions of the problem allow only transmitted waves traveling to the right after passing through the double barrier (i.e., $B_5 = 0$). Moreover, and without loss of generality, we can normalize the incident beam to unit amplitude (i.e., $A_1 = 1$).

Matching the spinor wavefunctions at the four boundaries defined by $|x| = a_-$ and $|x| = a_+ + a_-$ gives relations between $(A_\nu, B_\nu)$ in $\nu$-th region and those in the neighboring region. We prefer to express these relationships in terms of 2×2 transfer matrices between different regions, $\{M_n\}$, where $\binom{A_n}{B_n} = M_n \binom{A_{n+1}}{B_{n+1}}$. Finally, we obtain the full transfer matrix over the whole double barrier which can be written, in an obvious notation, as follows

$$\binom{1}{R} = \left(\prod_{n=1}^{4} M_n\right)\binom{T}{0} = M(E)\binom{T}{0}, \tag{9}$$

where $M(E) = M_1 M_2 M_3 M_4$ and we have set $R = B_1$ and $T = A_5$; $R$ and $T$ being the reflection and transmission amplitudes, respectively. We have assumed an incident wave from left normalized to unit amplitude (i.e., $A_1 = 1$ and $B_5 = 0$). The explicit form of the transfer matrices $M_n$ depends on the specific energy range. There are three such ranges for all $E > m$. These ranges are: (i) $m < E < V_-$, (ii) $V_- < E < V_+$ and (iii) $E > V_+$. Therefore, we end up with the full set of twelve transfer matrices given in the Appendix. Eq. (9) leads to

$$T(E) = 1/M_{11}(E), \quad R(E) = M_{21}(E)/M_{11}(E). \tag{10}$$

Time reversal invariance and the relevant conservation laws dictate that the transfer matrix $M(E)$ has a unit determinant along with the following symmetry properties $M_{11}(E) = M_{22}(E)^*$ and $M_{12}(E) = M_{21}(E)^*$. These can easily be checked using the explicit forms given in the Appendix. Thus, symmetry considerations impose strong conditions on the structure of the transfer matrix. Using these properties in Eq. (10) gives the expected flux conservation $|T|^2 + |R|^2 = 1$. Moreover, from Eq. (10) we see that full transmission or resonance transmission occurs at energies where the condition $|M_{11}(E)| = 1$ is satisfied (equivalently, $|M_{21}(E)| = 0$).

## III. RESULTS & DISCUSSION

The physical content of particle scattering through the double barrier depends on the energy of the incoming particle, which can assume any value larger than $mc^2$. In order to allow for super-criticality of the potential scattering we need to impose certain conditions on the heights of the potential barriers. Our study concentrates on Klein energy zones where full transmission can take place. The situation of interest to our study is for two Klein energy zones, which arises when $V_- > 2m$ and $V_+ - V_- > 2m$. As an example, we calculate the transmission coefficient as a function of energy for a given set of potential parameters. The result is shown in Fig. 2. In addition to the expected above-barrier full transmission for some values of energies larger than $V_+ + m$, one can clearly identify three sub-barrier regions where transmission resonances occur. These are:

i) The lower Klein energy zone ($m < E < V_- - m$): A region of seven resonances.
ii) The higher Klein energy zone ($V_- + m < E < V_+ - m$): A region of four resonances.
iii) The relativistic version of the conventional nonrelativistic double barrier transmission resonance ($V_+ - m < E < V_+ + m$): A region of four resonances.

It is also clear that resonances in the conventional energy zone are very sharp (or very narrow) whereas those in the two Klein energy zones are broad (or wide). This means that resonance states corresponding to the former decay much slower and have longer tunneling time than the latter. In Table 1, we list the resonance energies for this potential configuration to an accuracy of 10 decimal places. These were obtained as solution to the equation $M_{21}(E) = 0$. As further insight into the dynamics of this relativistic model we give an animation "a_m.mpg" of Fig. 2 as the distance between the two barriers, $2a_-$, varies from $2/m$ to $6/m$ [16]. The animation shows that:

i) The density of resonances in each of the three sub-barrier regions increases with $a_-$. That is, the energy separation between resonances decreases with $a_-$.
ii) As $a_-$ increases, resonance energies drop down (fall or dive) from the above-barrier region into the conventional resonance region then into the higher Klein energy zone.
iii) Additionally, as $a_-$ increases, resonance energies are created at the bottom of the spectrum (at $E \sim m$) then move up into the lower Klein energy zone.

We also give another animation "a_p.mpg" of Fig. 2 as the width of the barriers, $a_+$, varies from $1/m$ to $3/m$ [17]. The animation shows that all resonances get sharper with an increase in $a_+$ but the number of resonances in the conventional region does not change (i.e., the population density of resonances in this region is independent of $a_+$). We would like to mention a related recent work by Villalba and Gonzales-Arraga [18] who considered the resonant tunneling through a double square barrier and double cusp potentials. Our problem differs from that in [18] by the choice of an elevated floor of the potential well, which gives rise to two Klein energy zones of resonance. This potential design gave rise to a peculiar energy dependence of the transmission with three resonance regions; one is due to the conventional quantum tunneling and two others are due to Klein tunneling.

Finally, we note that the present work will not remain at this stage but will be followed by another. We plan to use the results obtained so far to deal with different issues related to transport properties in graphene. One of the main characteristics of Dirac fermions in graphene is the accuracy with which we can model their behavior by having extremely small mass (in fact, even massless). This implies that at any finite energy the model should be treated relativistically. This endows fermions in graphene with the ability to tunnel through a single potential barrier with probability one [11,19]. It is then natural to extend that analysis to our two-barrier problem case and investigate the basic features of such a system. However, we are reluctant to extend the present calculation (in the limit $m \to 0$) to the massless case without special care because it is well known that the massless and massive Dirac equations pertain to two completely different space-time symmetry groups; the conformal group and the Poincaré group, respectively [20]. Thus, special care must be taken to extend results from the massive to the massless case.

## ACKNOWLEDGEMENTS

This work is sponsored by the Saudi Center for Theoretical Physics (SCTP). Partial support by King Fahd University of Petroleum and Minerals is acknowledged.

# APPENDIX: TRANSFER MATRICES

If we define $a = a_+ + a_-$, $\sigma_0 = e^{ak_0}$, $\sigma_+ = e^{ak_+}$, and $\gamma_\pm = e^{a_-k_\pm}$ then in the first energy interval, $m < E < V_-$, the four transfer matrices at the boundaries $|x| = a_-$ and $|x| = a$ are given by

$$M_1 = \frac{1}{2}\begin{pmatrix} \sigma_0\sigma_+\left(\frac{1}{\alpha_0} - \beta_+\right) & \frac{\sigma_0}{\sigma_+}\left(\frac{1}{\alpha_0} + \beta_+\right) \\ -\frac{\sigma_+}{\sigma_0}\left(\frac{1}{\alpha_0} + \beta_+\right) & \frac{1}{\sigma_0\sigma_+}\left(\beta_+ - \frac{1}{\alpha_0}\right) \end{pmatrix}, \quad M_2 = \frac{1}{2}\begin{pmatrix} \frac{\gamma_-}{\gamma_+}\left(1 + \frac{\beta_-}{\beta_+}\right) & \frac{1}{\gamma_+\gamma_-}\left(1 - \frac{\beta_-}{\beta_+}\right) \\ \gamma_+\gamma_-\left(1 - \frac{\beta_-}{\beta_+}\right) & \frac{\gamma_+}{\gamma_-}\left(1 + \frac{\beta_-}{\beta_+}\right) \end{pmatrix},$$

$$M_3 = \frac{1}{2}\begin{pmatrix} \frac{\gamma_-}{\gamma_+}\left(1 + \frac{\beta_+}{\beta_-}\right) & \gamma_+\gamma_-\left(1 - \frac{\beta_+}{\beta_-}\right) \\ \frac{1}{\gamma_+\gamma_-}\left(1 - \frac{\beta_+}{\beta_-}\right) & \frac{\gamma_+}{\gamma_-}\left(1 + \frac{\beta_+}{\beta_-}\right) \end{pmatrix}, \quad M_4 = \frac{1}{2}\begin{pmatrix} \sigma_0\sigma_+\left(\alpha_0 - \frac{1}{\beta_+}\right) & -\frac{\sigma_+}{\sigma_0}\left(\alpha_0 + \frac{1}{\beta_+}\right) \\ \frac{\sigma_0}{\sigma_+}\left(\alpha_0 + \frac{1}{\beta_+}\right) & \frac{1}{\sigma_0\sigma_+}\left(\frac{1}{\beta_+} - \alpha_0\right) \end{pmatrix}.$$

(A1)

Then in the second energy range, $V_- < E < V_+$, we have

$$M_1 = \frac{1}{2}\begin{pmatrix} \sigma_0\sigma_+\left(\frac{1}{\alpha_0} - \beta_+\right) & \frac{\sigma_0}{\sigma_+}\left(\frac{1}{\alpha_0} + \beta_+\right) \\ -\frac{\sigma_+}{\sigma_0}\left(\frac{1}{\alpha_0} + \beta_+\right) & \frac{1}{\sigma_0\sigma_+}\left(\beta_+ - \frac{1}{\alpha_0}\right) \end{pmatrix}, \quad M_2 = \frac{1}{2}\begin{pmatrix} \frac{1}{\gamma_+\gamma_-}\left(\alpha_- - \frac{1}{\beta_+}\right) & -\frac{\gamma_-}{\gamma_+}\left(\alpha_- + \frac{1}{\beta_+}\right) \\ \frac{\gamma_+}{\gamma_-}\left(\alpha_- + \frac{1}{\beta_+}\right) & \gamma_+\gamma_-\left(\frac{1}{\beta_+} - \alpha_-\right) \end{pmatrix},$$

$$M_3 = \frac{1}{2}\begin{pmatrix} \frac{1}{\gamma_+\gamma_-}\left(\frac{1}{\alpha_-} - \beta_+\right) & \frac{\gamma_+}{\gamma_-}\left(\frac{1}{\alpha_-} + \beta_+\right) \\ -\frac{\gamma_-}{\gamma_+}\left(\frac{1}{\alpha_-} + \beta_+\right) & \gamma_+\gamma_-\left(\beta_+ - \frac{1}{\alpha_-}\right) \end{pmatrix}, \quad M_4 = \frac{1}{2}\begin{pmatrix} \sigma_0\sigma_+\left(\alpha_0 - \frac{1}{\beta_+}\right) & -\frac{\sigma_+}{\sigma_0}\left(\alpha_0 + \frac{1}{\beta_+}\right) \\ \frac{\sigma_0}{\sigma_+}\left(\alpha_0 + \frac{1}{\beta_+}\right) & \frac{1}{\sigma_0\sigma_+}\left(\frac{1}{\beta_+} - \alpha_0\right) \end{pmatrix}.$$

(A2)

Note that $M_1$ and $M_4$ have the same form as in (A1). Finally in the energy range, $E > V_+$, we obtain

$$M_1 = \frac{1}{2}\begin{pmatrix} \frac{\sigma_0}{\sigma_+}\left(1 + \frac{\alpha_+}{\alpha_0}\right) & \sigma_0\sigma_+\left(1 - \frac{\alpha_+}{\alpha_0}\right) \\ \frac{1}{\sigma_0\sigma_+}\left(1 - \frac{\alpha_+}{\alpha_0}\right) & \frac{\sigma_+}{\sigma_0}\left(1 + \frac{\alpha_+}{\alpha_0}\right) \end{pmatrix}, \quad M_2 = \frac{1}{2}\begin{pmatrix} \frac{\gamma_+}{\gamma_-}\left(1 + \frac{\alpha_-}{\alpha_+}\right) & \gamma_+\gamma_-\left(1 - \frac{\alpha_-}{\alpha_+}\right) \\ \frac{1}{\gamma_+\gamma_-}\left(1 - \frac{\alpha_-}{\alpha_+}\right) & \frac{\gamma_-}{\gamma_+}\left(1 + \frac{\alpha_-}{\alpha_+}\right) \end{pmatrix},$$

$$M_3 = \frac{1}{2}\begin{pmatrix} \frac{\gamma_+}{\gamma_-}\left(1 + \frac{\alpha_+}{\alpha_-}\right) & \frac{1}{\gamma_+\gamma_-}\left(1 - \frac{\alpha_+}{\alpha_-}\right) \\ \gamma_+\gamma_-\left(1 - \frac{\alpha_+}{\alpha_-}\right) & \frac{\gamma_-}{\gamma_+}\left(1 + \frac{\alpha_+}{\alpha_-}\right) \end{pmatrix}, \quad M_4 = \frac{1}{2}\begin{pmatrix} \frac{\sigma_0}{\sigma_+}\left(1 + \frac{\alpha_0}{\alpha_+}\right) & \frac{1}{\sigma_0\sigma_+}\left(1 - \frac{\alpha_0}{\alpha_+}\right) \\ \sigma_0\sigma_+\left(1 - \frac{\alpha_0}{\alpha_+}\right) & \frac{\sigma_+}{\sigma_0}\left(1 + \frac{\alpha_0}{\alpha_+}\right) \end{pmatrix}.$$

(A3)

**Table Caption:**

**Table 1:**
Transmission resonance energies (in units of $mc^2$) for the potential configuration associated with Fig. 2 ($V_+ = 8m$, $V_- = 4m$, $a_+ = 3/m$, $a_- = 2.5/m$). These values were obtained as solutions to the equation $M_{21}(E) = 0$.

**Table 1**

| Level | Lower Klein energy zone | Higher Klein energy zone | Conventional energy zone | Above-Barrier energy zone |
|---|---|---|---|---|
| 0 | 1.1913921248 | 5.1824247690 | 7.2022544582 | 9.1265979020 |
| 1 | 1.4523858967 | 5.5378483868 | 7.7033320458 | 9.4112532302 |
| 2 | 1.7708714661 | 6.1348174089 | 8.2103665633 | 9.4794638424 |
| 3 | 2.0643517404 | 6.7590893689 | 8.7007206203 | 9.7910774141 |
| 4 | 2.2185949080 | | | 10.1475989665 |
| 5 | 2.4744005714 | | | 10.3256144827 |
| 6 | 2.7966547987 | | | 10.5446769142 |
| 7 | | | | 10.9095057090 |

**Figure Captions**

**Fig. 1**
The potential configuration of the relativistic double barrier problem with $V_- > 2m$ and $V_+ > V_- + 2m$. Oscillatory solutions are in the grey regions, whereas exponential solutions are in the white regions. The oscillatory positive/negative energy solutions are located in the light/dark grey areas.

**Fig. 2**
The transmission coefficient as a function of energy for $V_+ = 8m$, $V_- = 4m$, $a_+ = 3/m$, and $2a_- = 5/m$. Evident are the three sub-barrier transmission-resonance regions. The lowest two are within the two Klein energy zones and the highest one with sharp resonances is bounded within the energy range $V_+ \pm m$.

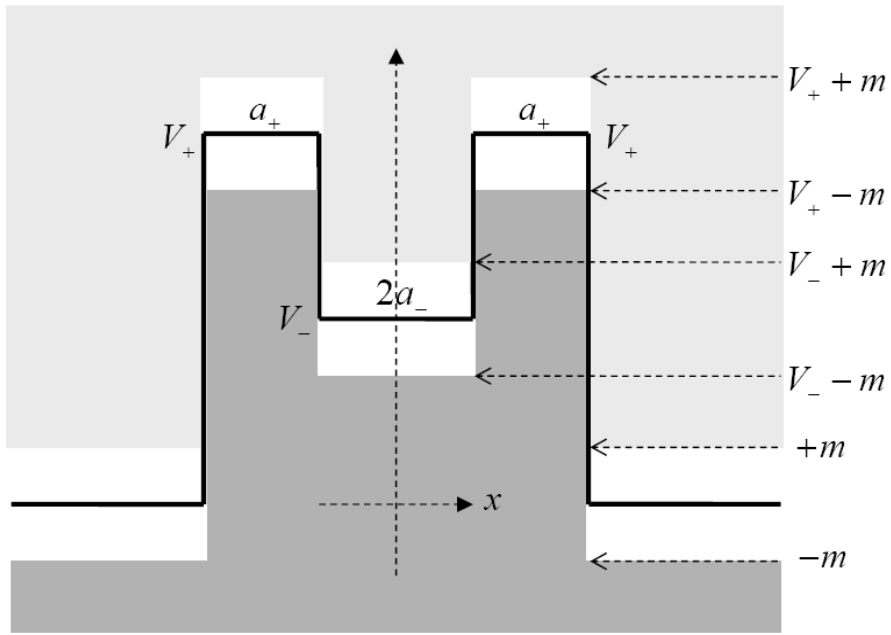

**Fig. 1**

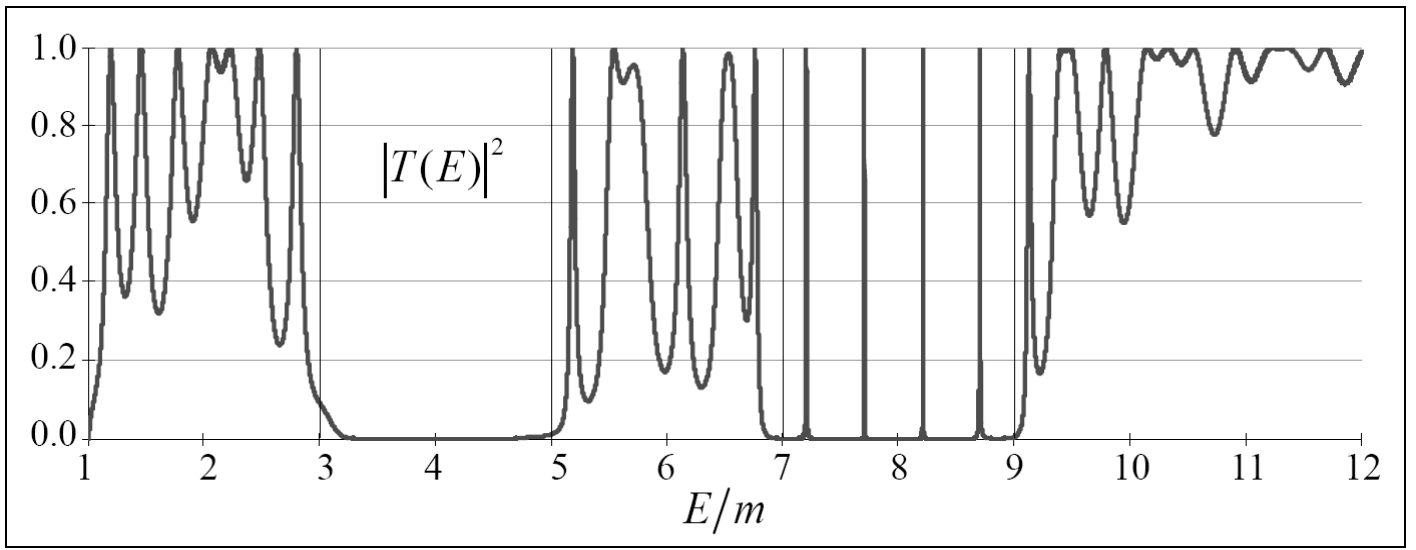

**Fig. 2**